\documentclass{aastex62}
\usepackage{epstopdf}
\usepackage{graphicx}
\usepackage[latin10]{inputenc}

\begin{document}
\title{Activity Complexes and A Prominent Poleward Surge During Solar Cycle 24}

\correspondingauthor{Jie Jiang}
\email{jiejiang@buaa.edu.cn}

\author{Zi-Fan Wang}
\affiliation{Key Laboratory of Solar Activity, National Astronomical Observatories, Chinese Academy of Sciences, Beijing 100101, China}
\affiliation{School of Astronomy and Space Science, University of Chinese Academy of Sciences, Beijing, China}

\author[0000-0001-5002-0577]{Jie Jiang}
\affiliation{School of Space and Environment, Beihang University, Beijing, China}
\affiliation{Key Laboratory of Space Environment Monitoring and Information Processing of MIIT, Beijing, China}

\author{Jie Zhang}
\affiliation{Department of Physics and Astronomy, George Mason University, Fairfax, VA 22030, USA}

\author{Jing-Xiu Wang}
\affiliation{School of Astronomy and Space Science, University of Chinese Academy of Sciences, Beijing, China}
\affiliation{Key Laboratory of Solar Activity, National Astronomical Observatories, Chinese Academy of Sciences, Beijing 100101, China}

\begin{abstract}
Long-lasting activity complexes (ACs), characterised as a series of closely located, continuously emerging solar active regions (ARs), are considered generating prominent poleward surges from observations.  The surges lead to significant variations of the polar field, which are important for the modulation of solar cycles.  We aim to study a prominent poleward surge during solar cycle 24 on the southern hemisphere, and analyse its originating ACs and the effect on the polar field evolution.  We automatically identify and characterize ARs based on synoptic magnetograms from the Solar Dynamic Observatory.  We assimilate these ARs with realistic magnetic configuration into a surface flux transport model, and simulate the creation and migration of the surge.  Our simulations well reproduce the characteristics of the surge and show that the prominent surge is mainly caused by the ARs belonging to two ACs during Carrington Rotations 2145-2159 (December 2013-January 2015).  The surge has a strong influence on the polar field evolution of the southern hemisphere during the latter half of cycle 24. Without the about one-year-long flux emergence in the form of ACs, the polar field around the cycle minimum would have remained at a low level and even reversed to the polarity at cycle 23 minimum.  Our study also shows that the long-lived unipolar regions due to the decay of the earlier emerging ARs cause an intrinsic difficulty of automatically identifying and precisely quantifying later emerging ARs in ACs. 

\end{abstract}

\section{Introduction} \label{sec:intro}



The spatial and temporal proximity of active regions (ARs) on the Sun is an important feature of large-scale solar magnetism \citep{1992ASPC...27...89V,1993SoPh..148...85H}.  A sequence of closely located, continuously emerging ARs is defined as a complex of activity \citep{1983ApJ...265.1056G}, or activity complex (AC).  This definition agrees with the concept of ``sunspot nests'' \citep[see][]{1986SoPh..105..237C}, which last for several rotations and contain over 30\% ARs in total.  ACs can also be associated with ``active longitudes'', which date back to \citet{1858MNRAS..19....1C}.  Consequently, ACs are foci of Sun's toroidal magnetic fields, and an indicator of solar cycle development \citep{2015ARep...59..228Y}.  The nesting tendency of ARs and physical interpretations are discussed in the review of \citet{2015LRSP...12....1V}.



The decay of ARs over the solar surface determines the evolution of the large-scale field, including the polar field.  According to the Babcock-Leighton (B-L) mechanism \citep{1961ApJ...133..572B,1964ApJ...140.1547L,1969ApJ...156....1L}, the surface large-scale field corresponds to the surface poloidal field of the dynamo loop and serves as the seed of the subsequent cycle \citep{2015Sci...347.1333C,2017SSRv..210..351W,2018ApJ...863..159J,2020LRSP...17....2P}.  The polar field is also the source of the fast solar wind \citep{2005Sci...308..519T} and the interplanetary field \citep{1995Sci...268.1007B,2010ApJ...709..301J}, which modulates the density of cosmic rays \citep{1999GeoRL..26..565C,2013LRSP...10....3P}.  Considering the continuous flux emergence in ACs, the decay of ACs is possible to introduce strong variations to the surface large-scale field, and is supposed to be a cause of short-term variation and perturbation of the interplanetary field.  Thus it is important to investigate the evolution of ARs within ACs and its influence on the large-scale field.


The migration of ARs on the surface can be described by the surface flux transport (SFT) model \citep[e.g.][]{1985AuJPh..38..999D,1989ApJ...347..529W,1998ApJ...501..866V,2002SoPh..209..287M,2014ApJ...791....5J}.  The SFT model utilizes a set of transport parameters based on observations and estimations to solve the magnetic induction equation on the surface, with the source of surface magnetic flux coming from emerging flux in ARs.  The SFT model is able to reproduce the polar field evolution.  Early SFT simulations use the bipolar magnetic region (BMR) approximation as the source term.  \citet{1989ApJ...347..529W} analysed the evolution of BMRs by comparing SFT simulations with observational longitudinally averaged surface field during cycle 21, and found how the trailing polarity flux of BMRs built up the polar field. \citet{2010ApJ...719..264C} used BMRs constructed from observations in the SFT model to reproduce the solar open flux and polar field of cycles 15-21.


Recently, more SFT simulations have begun to use real configurations instead of the BMR approximation as source terms \citep[e.g.][]{2015SoPh..290.3189Y,2017A&A...604A...8V,2018ApJ...863..116W,2019ApJ...871...16J}.  Using real configuration of ARs is more precise in terms of polar field influence than using BMRs constructed from AR parameters, especially for more complex ones like $\delta$-type ARs \citep{2019ApJ...871...16J}.  The final dipole moment could deviate from or even be opposite to that determined by tilt angle and other AR parameters.  Therefore, using real configurations of ARs is necessary as we aim to simulate ARs emerging during a few Carrington Rotations (CRs) in detail.


The poleward migration of flux is not uniform in time, but usually in the form of poleward surges (or poleward flux streams, plumes) seen on magnetic butterfly diagrams, i.e., longitudinally averaged radial magnetic field at the photosphere \citep{1981SoPh...74..131H}.  Poleward surges' characteristics are related to originating ARs, whose axes have tilts respect to the east-west direction.  Typically, a tilted bipolar AR generates a poleward surge of the trailing polarity. This results from the latitudinal separation of the two polarities because of the tilt angle, which leads to a more diffusive, less distinguishable poleward migration of the leading polarity \citep{2002SoPh..209..287M,2015SoPh..290.3189Y,2015ApJ...798..114S,2019ApJ...871...16J}.  Surges are more concentrated if originating from higher latitudes, and more diffusive if originating from lower latitudes.  The concentrated surges correspond to a short-term perturbation in the polar field by ARs.  On the other hand, ARs' influences on the final polar field, i.e., the polar field at cycle minimum, decrease exponentially as their emergence latitudes increase, as revealed by the SFT simulations of \citet{2014ApJ...791....5J,2018ApJ...863..116W}, and by a mathematical explanation of \citet{2020arXiv200902299P}.  Hence, a surge with long-lasting influence to the solar cycle development should originate from lower latitudes, especially with concentrated flux emergence, e.g., ACs.

Meanwhile, observations suggest the relation among ACs, the generated poleward surges, and the polar field reversal \citep{2016STP.....2a...3M,2019SoPh..294...21M}.  ARs with different tilt angles in ACs may form large regions of single polarity as a result of flux cancellation between them, and create poleward surges \citep{2008ApJ...686.1432G}.  Comparison of polar field reversal of cycles 21-24 shows that ACs with larger area and longer lifetime are associated with stronger poleward surges and more violent polar field reversal \citep{2017SSRv..210...77P}.  This, however, was done by stacking synoptic maps, as \citet{1983ApJ...265.1056G} did, and by measuring observational magnetic flux at different latitudes, but no SFT simulations were applied.  It is needed to tell whether such long-lasting ACs that are able to cause violent polar field variation can have significant influence on the final polar field as well.  The relationship between poleward surges by ACs and their polar field influence should be evaluated in the data-driven SFT model to tell the exact long-term influence of ACs.

\begin{figure}
\gridline{\fig{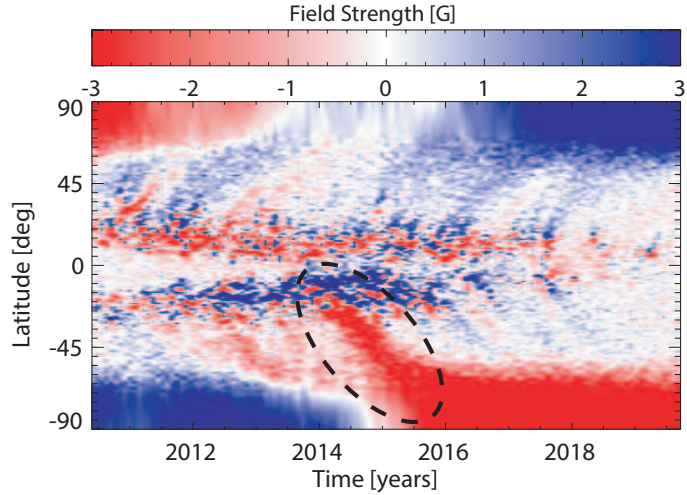}{0.5\textwidth}{}          }
\caption{Magnetic butterfly diagram of cycle 24 generated from longitudinally averaged SDO/HMI synoptic maps described in Subsection \ref{sibsec:identifyARs}.  The poleward surge studied in the article is marked with the black dashed ellipse in the figure.\label{fig:magbfly}}
\end{figure}

To learn the contribution of ACs to poleward surges and the final polar field, we simulate the production and migration of the most prominent poleward surge on the southern hemisphere during cycle 24.  As shown in Figure \ref{fig:magbfly}, the surge of interest covers the time period of CRs 2145-2159 (DEC 2013-JAN 2015), carrying a large quantity of negative flux that reversed the polar field from positive to negative and further strengthened the field to -4 G on average.  The surge of interest originates from long-lasting ACs, which will be shown in Subsection \ref{subsec:identifyResults}.  We simulate the overall development of the surge by assimilating the ARs with their real configurations during CRs 2145-2159 into SFT simulations, and compare them with observations.  Our simulation is able to reproduce the features of the surge, especially its dominant influence on the southern polar field.  We present the simulation of a fraction of the ACs in detail, discussing the continuous flux emergence and cancellation occurring in ACs, in order to examine and discuss the associated intrinsic difficulty of automatic AR identification in ACs.

The article is organized as follows.  In Section \ref{sec:ars} we describe the AR identification method based on synoptic maps.  We then show the identified ARs and describe the ACs.  In Section \ref{sec:sft} we introduce the data driven SFT model that we use, and the assimilation technique of identified ARs.  In Section \ref{sec:results} we present simulation results for the surge.  We discuss and conclude in Section \ref{sec:conclusion}.

\section{ARs during CRs 2145-2159} \label{sec:ars}

\subsection{Method of identifying ARs} \label{sibsec:identifyARs}

To obtain the input source term for the SFT model, an AR identification method is needed.  Up to now, identification of ARs for the source term of the SFT model has been done by smoothing the magnetograms and applying a threshold, as done by \citet{2015SoPh..290.3189Y,2017A&A...604A...8V,2017A&A...607A..76W,2018ApJ...863..116W}.  The threshold is determined by trial and optimization methods over a given cycle.  However, as we aim to examine the properties of ACs within a shorter period of time than the whole cycle, where flux of different strengths and concentrations is mixed as a result of continuous emergence, one threshold is not sufficient to obtain the needed AR properties.  Instead, we apply an identification method with the ability of morphological analysis to adapt to the complex environment in ACs.

AR data are obtained from synoptic maps of radial magnetic field component from Helioseismic and Magnetic Imager of Solar Dynamics Observatory (SDO/HMI) \citep{2012SoPh..275..229S}. The radial field component is derived from HMI line-of-sight 720 sec-cadence magnetograms by dividing cosine latitude.  The synoptic maps have a size of 3600$\times$1440 pixels, with equal space in longitude and equal space in sine-latitude.  We choose the polar field interpolated version created by the method of \citet{2011SoPh..270....9S,2018arXiv180104265S}. 

We use the AR identification method of \citet{2010ApJ...723.1006Z} to obtain the ARs needed for the SFT model input.  The method is originally created for MDI magnetograms, and is adapted to the HMI data in this study.  In order to carry out the  morphological analysis, the method needs four different control parameters: kernel threshold, erosion size, growth threshold, and dilation size.  The method selects regions with strength larger than the kernel threshold and area larger than the erosion size as kernels.  The kernels are grown up to the growth threshold.  Grown regions within the dilation size are joined together to connect different parts of a single AR.  How to choose this set of four parameters depends on the conditions of synoptic maps and our purpose.

For the majority of the ARs, we use the following set of control parameters: kernel threshold 250 G, erosion size 10 Mm, growth threshold 45 G, and dilation size 10 Mm.  The set of parameters has been tested with different trials, and is found to successfully recover most ARs; it is referred to as the standard set.  A lower kernel threshold or a lower erosion size will keep smaller magnetic structures during the identification, resulting in more ARs identified.  A lower growth threshold will result in larger areas of identified ARs.  Here we show the extracted parameters of NOAA AR12222 on CR2157 as an example, as shown in Table \ref{tab:12222Changeparams}.  ARs' area and flux are calculated by summing the values over all AR pixels.  The longitude and latitude of ARs are determined by the geometric center of valid AR pixels weighted by unsigned fluxes.  Tilt angles are calculated from the weighted centers of two polarities.  We note that some identified regions are not well-balanced in positive and negative flux.  We select those ARs that are fairly balanced in magnetic flux, which means that the stronger polarity is less than three times that of the weaker polarity, such as AR12222 listed here.  This standard has also been applied to previous SFT simulations \citep[see][]{2017A&A...604A...8V}.

The first row of Table \ref{tab:12222Changeparams} shows the physical quantities of AR12222 obtained based on the standard set of control parameters. The second to the fourth rows correspond to the variations of the control parameters by changing the kernel threshold to 225 G, the growth threshold to 35 G, and the erosion and dilation sizes to 30 Mm, respectively. As shown, the properties of AR12222 are well kept despite the changes of control parameters.  In practice, we accept a kernel threshold between 225 G and 250 G, an erosion and dilation size between 10 Mm and 20 Mm, and a growth threshold between 40 G and 50 G.

\begin{deluxetable*}{cccccccc}[b!]
\tablecaption{Parameters of AR12222 identified by the standard set of control parameters and the corresponding variations \label{tab:12222Changeparams}}
\tablecolumns{7}
\tablenum{1}
\tablewidth{0pt}
\tablehead{
\colhead{Identification parameters} &
\colhead{Latitude} &
\colhead{Longitude} &
\colhead{Area} &
\colhead{Positive flux} &
\colhead{Negative flux} &
\colhead{Tilt} &
\\
\colhead{} & \colhead{(degree)} & \colhead{(degree)} &
\colhead{($\mu$Hem)\tablenotemark{a}} & \colhead{(10$^{20}$Mx)} & \colhead{(10$^{20}$Mx)} & \colhead{(degree)}
}
\startdata
   Standard set    &  -20.2  &     82.6   &   3021.9    &  183.9   &   -138.6    &    10.2      \\
   Decreased kernel   &   -20.2   &    82.6  &   3021.9  &   183.9 &    -138.6   &    10.2    \\
   Decreased growth  &   -20.2   &    82.6  &  3289.3   &   185.0  &   -141.3   &    10.2   \\
   Increased erosion/dilation  &   -20.2   &    82.6  &   3021.9    &  183.9  &   -138.6   &    10.2   \\
      Special set    &  -20.5  &     83.5    &       4968.4    &  194.4  &   -170.3   &    14.9           \\
\enddata
\tablenotetext{a}{Millionths of the solar hemisphere.}
\end{deluxetable*}

\begin{figure}
\gridline{\fig{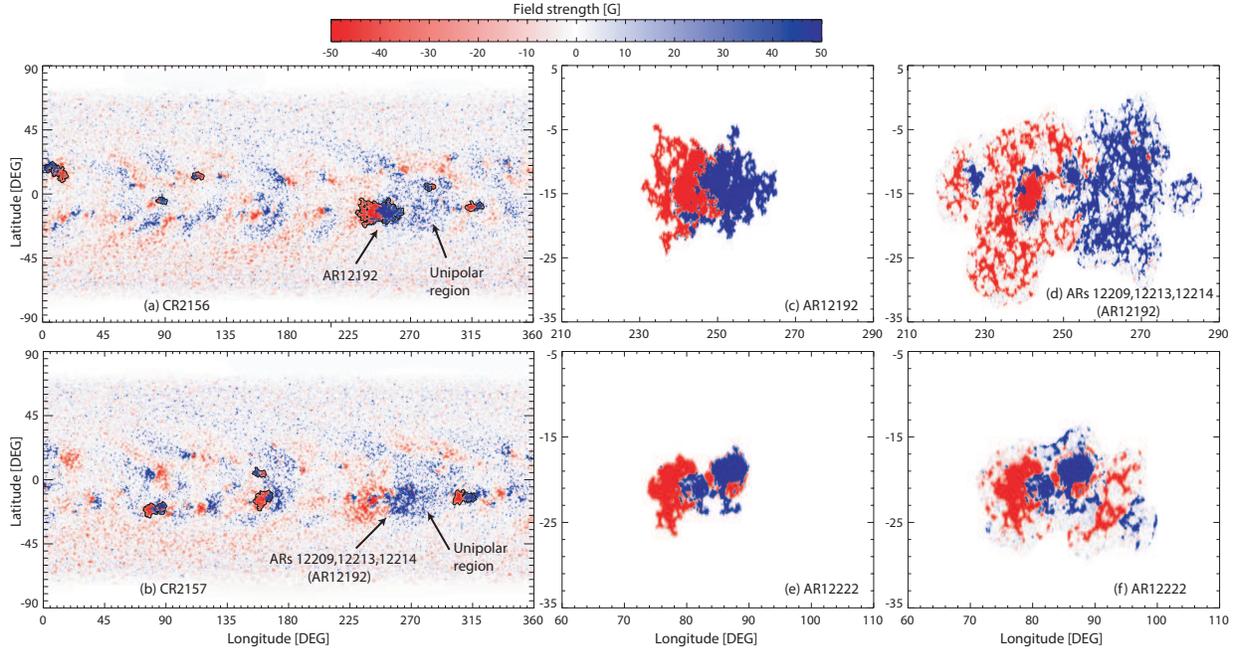}{0.9\textwidth}{}
 }
\caption{Identified ARs on synoptic maps.  The left column shows synoptic maps of CR2156 (a), and CR2157 (b), with the ARs (enclosed by black curves) identified by the automatic identification method using the standard set of control parameters.  AR12192 and the unipolar region in proximity are indicated with black arrows.  The middle and right columns show AR12192 identified using the standard set of control parameters on CR2156 (c), and identified using the special set of control parameters on CR2157 (d), as well as AR12222 identified using the standard set of control parameters on CR2157 (e), and identified using the special set of control parameters on CR2157 (f).\label{fig:identifyARs}}
\end{figure}

During the development of ACs, the identification of ARs within ACs can be problematic due to the extensive flux emergence and cancellation.  Some ARs are proximate to the unipolar regions generated by previous ACs, and may also be mixed with new flux emerging near or even within the considered AR.  In this case, the AR is composed of a mixture of stronger and weaker, more concentrated and more diffusive flux, and resides in a large area of non-zero background, which causes an intrinsic difficulty for automatic AR identification techniques with a given set of control parameters.  The AR with the largest area and longest time of occurrence during the time considered, labelled as NOAA AR12192, is such an example.  As shown in the synoptic maps in Figure \ref{fig:identifyARs}(a) and \ref{fig:identifyARs}(b), AR12192 is near a large unipolar region (in blue color), which originates from previous ARs in an AC.  We will simulate the formation of this unipolar region in Subsection \ref{subsec:differentARs}.  The standard set of control parameters is able to identify AR12192 on CR2156.  On CR2157, the recurring AR12192, relabelled as ARs 12209, 12213, and 12214, is mixed with new flux emergence near and within it.  We find that the standard set of control parameters is unable to identify AR12192 on CR2157.  Consequently we have to choose a different set of control parameters: kernel threshold 100 G, erosion size 10 Mm, growth threshold 20 G, and dilation size 60 Mm, to identify the more diffusive AR12192.  We refer to this set of control parameters as the special set.  As shown in Figure \ref{fig:identifyARs}(c) and \ref{fig:identifyARs}(d), the two configurations of identified AR12192 differ vastly.  The two configurations of AR12192 possess different parameters, presented in Table \ref{tab:12192params}.  During CRs 2156-2157, AR12192 shows an increase in area and a decrease in flux, and a notable change in tilt angle.  Our SFT simulation tests show that the evolution of AR12192 during CRs 2156-2157 cannot be achieved by SFT transport terms only.  This implies new flux emergences, as flux emergence is important in changing the properties of AR12192 \citep{2017ApJ...840..100M}.  Such new emergence superposes with existing flux, and cancels with existing opposite polarities.  Thus it is expected that the observed evolution of the AR during CRs 2156-2157 is different from the evolution solely deduced from the previous occurrence of the AR on CR2156.  Therefore, we choose the special set of control parameters for processing AR12192 on CR2157 for the following simulations.

The two sets of control parameters produce different results not only for AR12192, but for other ARs as well.  We show the difference for AR12222 as an example, in the last row of Table \ref{tab:12222Changeparams}.  As shown in Figure \ref{fig:identifyARs}(e) and \ref{fig:identifyARs}(f), when the special set of control parameters is applied, AR12222 shows a growth in size and flux, as well as a shift in configuration, for it is connected with other magnetic structures.  We note that the parameter set for AR12192 is an extreme case, and we keep the standard set of control parameters for all other ARs.

\begin{deluxetable*}{ccccccccc}[b!]
\tablecaption{Parameters of AR12192 identified on CR2156 and CR2157 \label{tab:12192params}}
\tablecolumns{8}
\tablenum{2}
\tablewidth{0pt}
\tablehead{
\colhead{Carrington Rotation} &
\colhead{Identification parameters} &
\colhead{Latitude} &
\colhead{Longitude} &
\colhead{Area} &
\colhead{Positive flux} &
\colhead{Negative flux} &
\colhead{Tilt} &
\\
\colhead{} &\colhead{} & \colhead{(degree)} & \colhead{(degree)} &
\colhead{($\mu$Hem)} & \colhead{(10$^{20}$Mx)} & \colhead{(10$^{20}$Mx)} & \colhead{(degree)}
}
\startdata
       2156   &  Standard set  &    -13.2   &   246.1  &  12779.4 &     631.2 &    -698.2    &    0.9    \\
       2157   &  Special set  &   -15.0   &   249.9   & 21834.3  &    546.0  &   -523.3   &    5.7     \\
\enddata
\end{deluxetable*}



\subsection{Identification results for ARs} \label{subsec:identifyResults}



\begin{figure}
\gridline{\fig{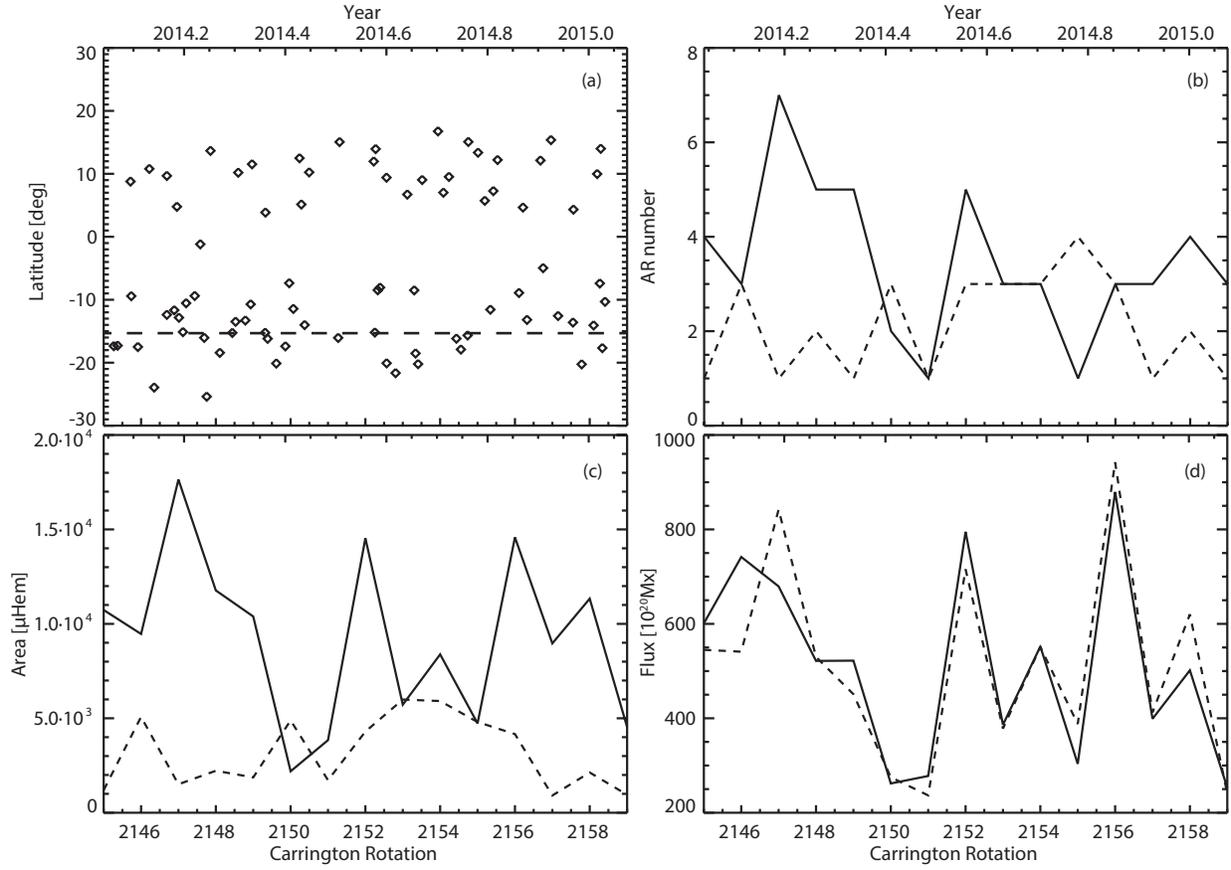}{0.9\textwidth}{}}

\caption{Identified ARs' parameters.  (a) Time-latitude diagram of identified ARs with each diamond representing one AR, with the dashed horizontal line at -15$^{\circ}$ latitude indicating the median of latitudes of ARs on the southern hemisphere; (b) Number of identified ARs in each hemisphere on each CR, with solid line indicating southern hemisphere and dashed line indicating northern hemisphere; (c) Total area of identified ARs on each CR, with solid line indicating southern hemisphere and dashed line indicating northern hemisphere;  (d) Total flux of identified ARs on each CR, with solid (dashed) line indicating negative (positive) flux.\label{fig:crotNar}}
\end{figure}

Utilizing the identification methods described above, we identified 84 ARs in total.  The time-latitude distribution is displayed in Figure \ref{fig:crotNar}(a).  As shown, all identified ARs reside within $\pm$30$^{\circ}$ latitude in both hemispheres, with the majority lying between -10$^{\circ}$ to -20$^{\circ}$.  The emergence latitudes of ARs do not follow a clear trend, since a relatively short time of period is considered here.  The majority of emerging latitudes are lower than the source of the surge studied by \citet{2015SoPh..290.3189Y}, which covers +20$^{\circ}$ to +40$^{\circ}$.  We expect the surge we study to produce large final polar field influence.

During CRs 2145-2159, more ARs are identified in the southern hemisphere than in the northern hemisphere, in terms of region counts and total areas.  Figure \ref{fig:crotNar}(b) and \ref{fig:crotNar}(c) show the numbers and total areas of ARs identified in both hemispheres on each CR, respectively.  The difference of areas between two hemispheres is larger than that of region numbers, indicating that the ARs of the southern hemisphere are also generally larger than regions of the northern hemisphere.

The imbalance of flux in two polarities is common among the identified ARs.  Figure \ref{fig:crotNar}(d) shows the total identified positive and negative flux of each CR.  The difference between two signs of polarities differs from rotation to rotation.  We note that unbalanced large ARs affect the result of our SFT model simulations.

\begin{figure}
\gridline{\fig{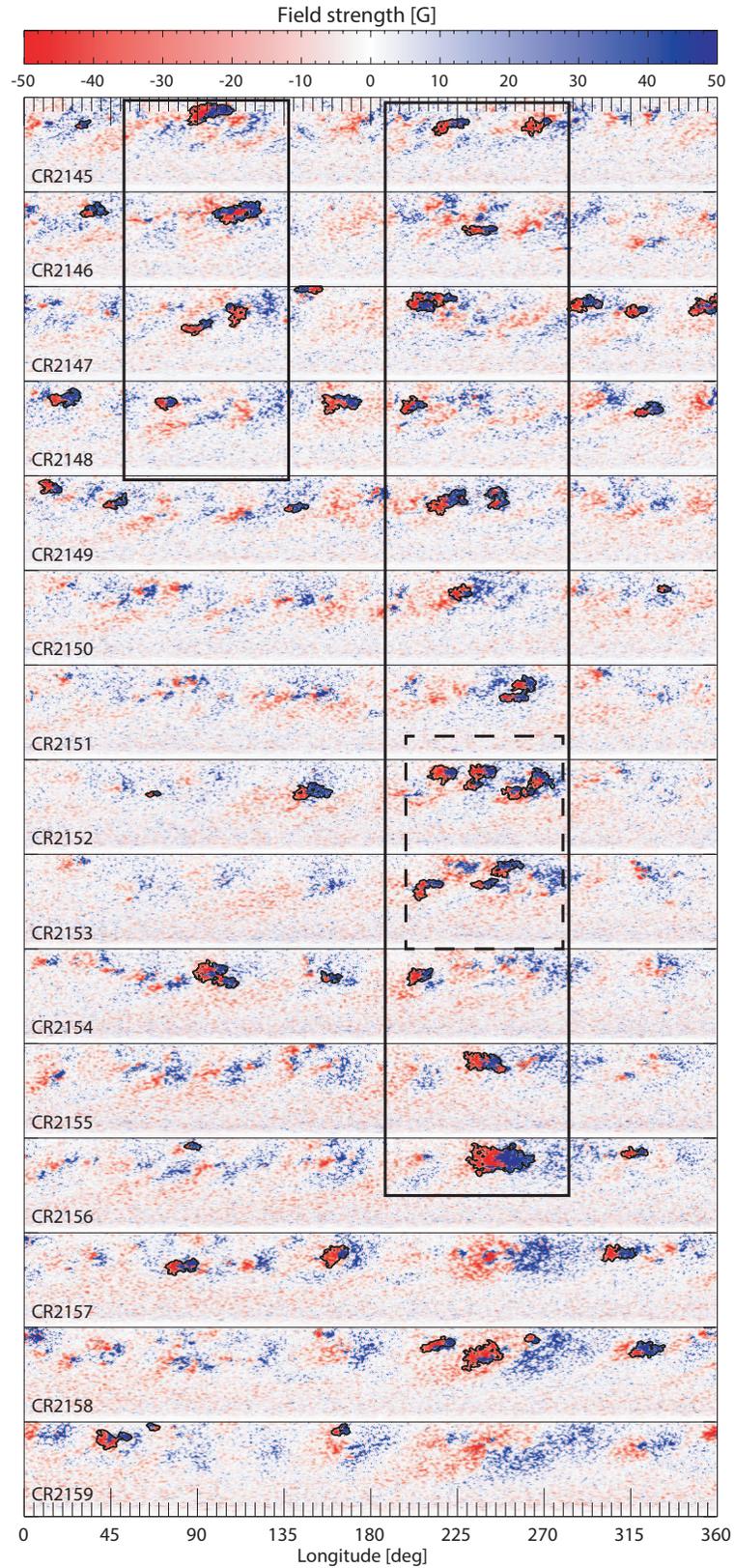}{0.55\textwidth}{}}
\caption{Stack plot of magnetograms during CRs 2145-2159 of the southern hemisphere, with the identified ARs outlined by the black curves.  Each magnetogram is displayed in equal sine-latitude in the latitudinal part.  Black solid squares mark the ARs that can be regarded as members of ACs.  The black dashed square marks the ARs studied in Subsection \ref{subsec:differentARs}.  AR12192 identified by the special set of control parameters on CR2157 is not marked here.\label{fig:arstack}}
\end{figure}

Among all identified ARs, some ARs in the southern hemisphere fit the concept of ACs.  We identify ACs according to standards similar to that of \citet{1983ApJ...265.1056G,2017SSRv..210...77P}.  By stacking those identified regions in the southern hemisphere rotation by rotation in Figure \ref{fig:arstack}, we see that ARs between 180$^{\circ}$ and 270$^{\circ}$ longitudes belong to a long lasting AC.   This major AC exists from CR2145 to CR2156, and ends as the AR12192 emerges; AR 12192 is the largest AR in area and the most abundant in flux in solar cycle 24.  This AC consists of 18 regions, over 30\% of all ARs in the southern hemisphere.  From its rotation-to-rotation development, we can clearly observe the formation of unipolar magnetic regions, especially the region that exists for several rotations near AR12192.  The formation of such regions is a superposition of flux from several previous ARs emerging around. After one CR from its major flux emergence, the diffusive AR12192 is mixed with the unipolar region, as well as new flux emergence.   From this perspective we consider AR12192 a member of the AC as well.  Besides the most long-lasting AC mentioned above, the ARs between 50$^{\circ}$ and 135$^{\circ}$ during CRs 2145-2149 can also be identified as an AC, or a nest.  The flux from these ARs also superposes during evolution, forming unipolar regions.  In total, approximately 50\% of all identified ARs are associated with the two ACs.



\section{SFT model with data driven source} \label{sec:sft}

\subsection{Model description} \label{subsec:sftDescribe}

The SFT model is to solve the radial component of the magnetic induction equation with given differential rotation, meridional flow and supergranular diffusion at the solar surface to get the temporal evolution of radial magnetic field component.  The SFT simulation code is based on \citet{2004A&A...426.1075B}.  The code has 360$\times$180 spatial resolution and a time interval of one day.  The spatial component is expanded in terms of spherical harmonics up to the order of 63, in which case the resolution corresponds to the size of supergranulation.  The temporal component is solved with the 4th-order Runge-Kutta method.


The differential rotation and meridional flow are determined empirically from historic observations.  We adopt the differential rotation profile of \citet{1983ApJ...270..288S}, and the meridional flow of \citet{1998ApJ...501..866V}.  The meridional flow speed is set to 11 ms$^{-1}$.  The supergranular diffusivity is set to 500 km$^2$s$^{-1}$.  The model details are described in \citet{2014ApJ...791....5J}.


Besides aforementioned time-independent transport processes, inflows towards activity belts are another important process that affects the result of SFT simulations \citep{2002ApJ...570..855H,2004ApJ...603..776Z,2004SoPh..224..217G,2010ApJ...717..597J,2012A&A...548A..57C}.  Inflows affect flux cancellation and the latitudinal separation of polarities of ARs \citep{2016A&A...586A..73M}.  For simulations that implement BMRs into the SFT code, the effect of inflows can be simplified as a factor that decreases all tilt angles of BMRs \citep{2015ApJ...808L..28J}.  For ARs with real configurations, the effect of inflows would be more complicated.  At present we do not include inflows in our simulations.  To compensate the effect of inflows, we set the supergranular diffusivity to 500 km$^2$s$^{-1}$, larger than the diffusivity of \citet{2014ApJ...791....5J}.  Still, inflows are expected to take part in the evolution of ACs, as ACs are characterized as a large amount of flux emergence and interaction in activity belts.  The exact effect of inflows remains to be discussed.

\subsection{Source term} \label{subsec:sourceTerm}


The ARs identified in Section \ref{sec:ars} are assimilated into the simulation.  We first balance the total flux of each AR by enlarging the flux of the weaker polarity on each pixel by a same ratio to balance positive and negative flux.  This balancing technique is identical to that of \citet{2019ApJ...871...16J}.  Then we convert the ARs from equal sine latitude to equal latitude, and rescale them to the resolution of the code.  The insertion of ARs is done by replacing the corresponding pixel in the simulation with new values of the ARs.  Considering how synoptic maps are generated, the day of inserting each AR into the simulation is determined by the time when it crosses central meridian.

Theoretically, ARs are intended to be assimilated into the SFT model when they reach the decay phase.  ACs are characterized as intense flux emergence and cancellation, so decaying ARs within ACs may be mixed with new emerging flux and old flux from previous ARs.  The observed flux is a superposition of flux from different emerging sources at different stages of evolution, so it is intrinsically hard to determine the exact decaying phase of ARs in an AC.  As shown in Subsection \ref{sibsec:identifyARs}, the recurring AR12192 on CR2157 has fairly different parameters and configuration from that of AR12192 on CR2156, possibly caused by flux emergence.  Hence, we use the configuration of AR12192 on CR2157, while its previous occurrence is not assimilated.  Such issue is a direct result of the identification problem in ACs described in Subsection \ref{sibsec:identifyARs}, and remains a possible problem for assimilation studies using real ARs.

\section{Results} \label{sec:results}

\subsection{The overall properties of the simulated poleward surge} \label{subsec:surge}

In order to simulate the poleward surge of interest, we first assimilate all identified ARs during CRs 2145-2159 (DEC 2013-JAN 2015).  We use the synoptic map of CR2144 as the initial magnetic field configuration, after converting it to equal latitude along the y-axis and rescaling it to the resolution of the code.  The simulation ends at CR2220 (AUG 2019), which was the latest CR with HMI synoptic maps available when the simulation was conducted.

\begin{figure}
\gridline{\fig{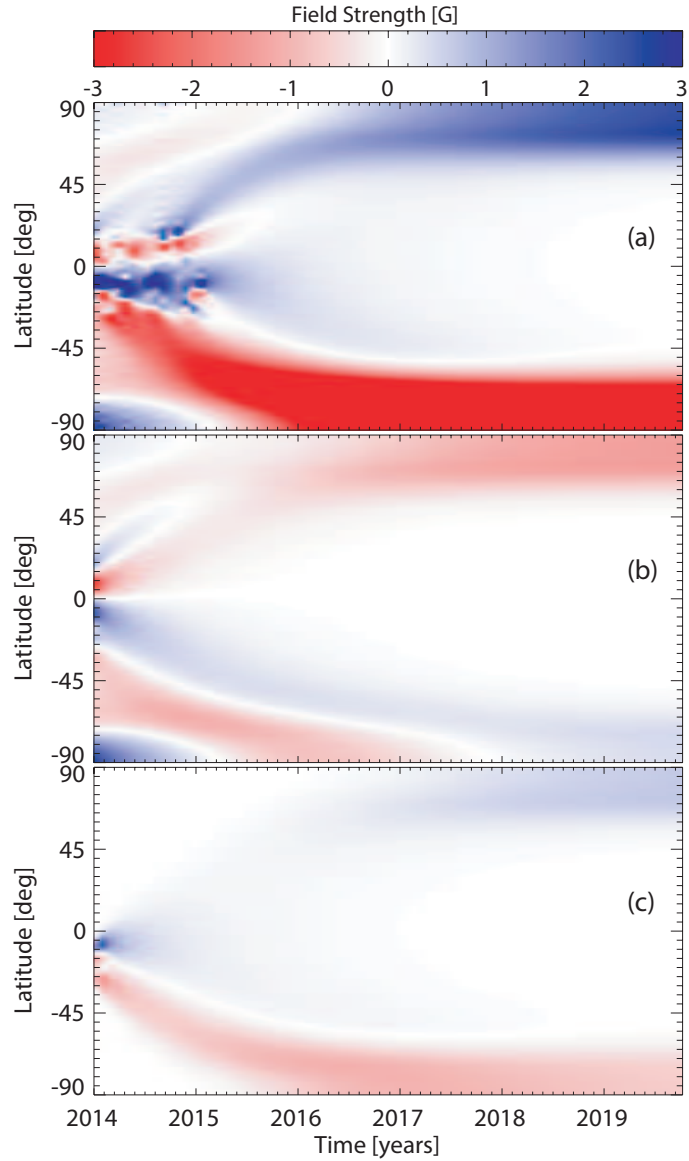}{0.5\textwidth} {}         }
\caption{Simulated magnetic butterfly diagrams for the simulation starting at CR2144 with ARs during CRs 2145-2159 assimilated (a), the simulation starting at CR2144 without ARs assimilated (b), and the simulation with no starting field and with ARs listed in Table \ref{tab:acarlist} assimilated (c).\label{fig:simMagbfly}}
\end{figure}

We generate the magnetic butterfly diagram for the whole simulated time range to display the simulated surge.  We also demonstrate a butterfly diagram of one simulation without the identified ARs, solving the evolution of the initial field only.  They are shown in Figure \ref{fig:simMagbfly}(a) and \ref{fig:simMagbfly}(b), respectively.  By comparing the diagrams we can show that the surge is primarily generated by the ARs in the time period.  Without such a surge and the originating ARs emerging during the year 2014, the polar field reversal would not have been achieved.  Instead it would remain at a low level, and even reverse to previous polarities.  The validation of our SFT simulation results is presented in the Appendix \ref{sec:validateSFT}.  In the following we present quantitative comparisons between the simulated results and observations.

\begin{figure}
\gridline{\fig{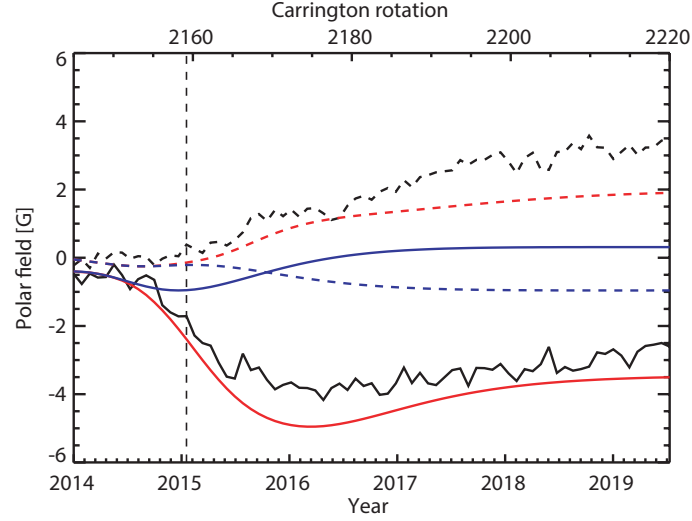}{0.5\textwidth}{}}
\caption{Time evolution of polar field.  Solid lines indicate the southern hemisphere, and dashed lines indicate the northern hemisphere.  Black lines are observational data.  Red lines are results of the simulation with initial field CR2144 and with ARs during CRs 2145-2159 assimilated.  Blue lines are results of the simulation with initial field CR2144 and without ARs assimilated.  The vertical black dashed line near year 2015 indicates CR 2159, the last CR with ARs assimilated.\label{fig:timePfAll}}
\end{figure}

Polar fields are obtained by averaging from 60$^{\circ}$ to 75$^{\circ}$ latitudes for both hemispheres.  Its evolution during CRs 2145-2220 is shown in Figure \ref{fig:timePfAll}.  Observations (black lines) show that from the middle of 2014 to the end of 2015, the southern polar field quickly rises to maximum from near zero, and then decreases gradually as the flux of the leading polarity from the originating ARs begins reaching the south pole, while the northern polar field rises more steadily and continuously.  The simulated polar field of the southern hemisphere (red solid line) is close to observations during the majority of the latter half of cycle 24, though ARs after CR2159 are not assimilated into the simulation.  This indicates that the rise to maximum and decrease from maximum of the southern polar field is mostly determined by the ARs assimilated, while the ARs after CR2159 do not have a significant effect on the southern polar field.  This confirms the importance of the ARs during CRs 2145-2159 to the final polar field, hence to the long-term development of the solar cycle.  According to \citet{2018ApJ...863..159J,2018JASTP.176...34J}, there are large ARs with abnormal tilt close to the equator during the years 2016 and 2017, weakening the contribution of ARs after CR2159 to the final polar field.  As a result, the final polar field at cycle 24 minimum is largely determined by the ARs during CRs 2145-2159.  Without ARs during the time period (blue lines), the final polar field would be extremely weak, leading to a possible next Maunder minimum. These ARs keep this from happening, and build up the final polar field, which has a similar strength to that of cycle 23.  Based on such strength of polar fields, we expect a moderate cycle 25.

\begin{figure}
\gridline{\fig{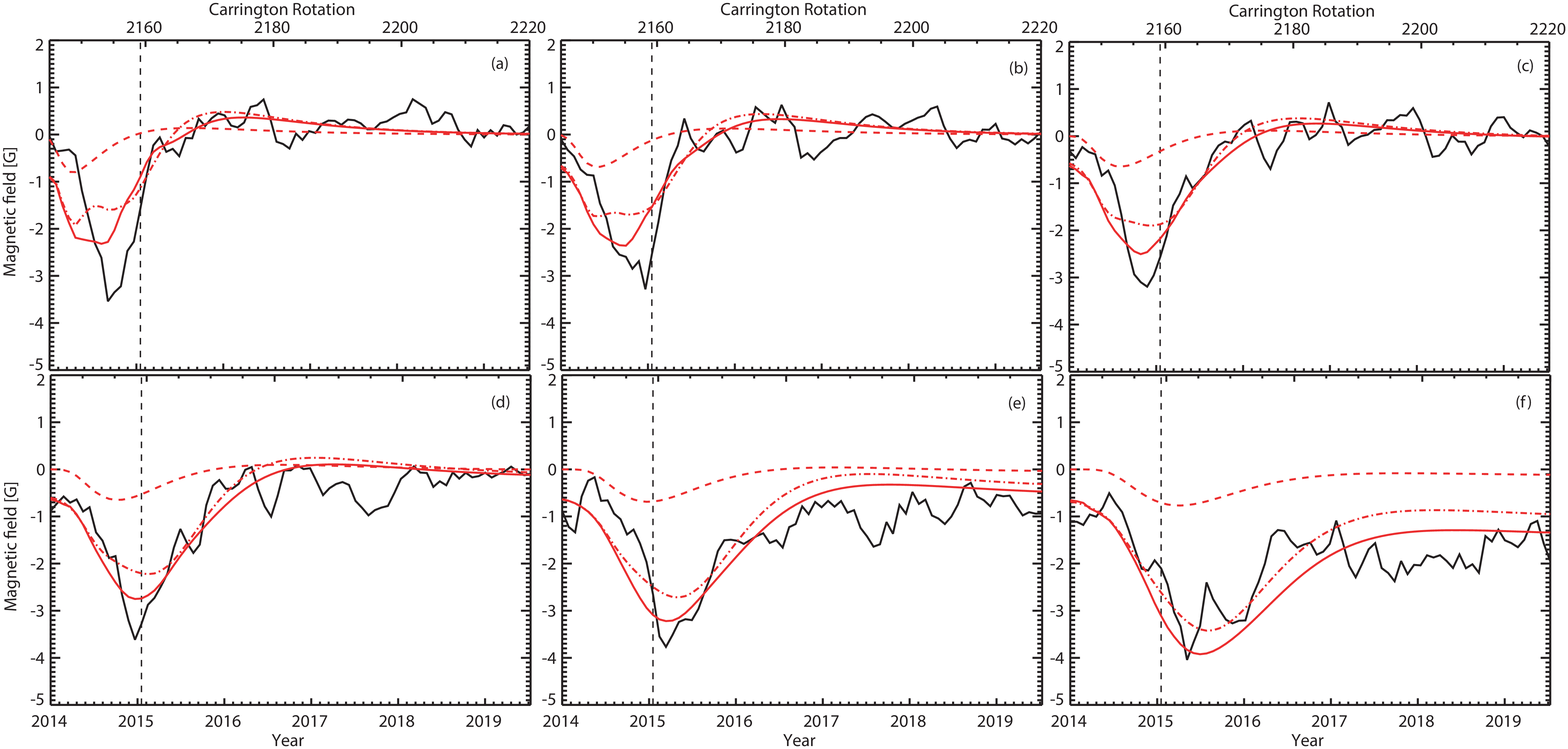}{0.9\textwidth}{}
          }
\caption{Comparisons of longitudinally averaged magnetic field on different latitudes of the southern hemisphere, obtained by slicing magnetic butterfly diagram on 35$^{\circ}$ (a), 40$^{\circ}$ (b), 45$^{\circ}$ (c), 50$^{\circ}$ (d), 55$^{\circ}$ (e), and 60$^{\circ}$ (f).  The black solid curves are observational, the red solid curves show the simulation including all ARs, the red dashed curves show the simulation including ARs listed in Table \ref{tab:acarlist}, and the red dash-dotted curves show the simulation including all ARs in ACs described in Subsection \ref{subsec:identifyResults}.  All curves are smoothed with a width of 3 rotations. The vertical black dashed lines indicate CR2159, the last CR with ARs assimilated.\label{fig:surgeLatitude}}
\end{figure}

The surge we consider is stronger and more influential than other poleward flux migrations in the latter half of cycle 24.  We show the observed and simulated field strength at different latitudes in Figure \ref{fig:surgeLatitude}.  As shown by the black curves obtained from observational data, the surge maintains its width of approximately 0.7-1.0 yr until it reaches high latitudes, that is, -55$^{\circ}$ (Figure \ref{fig:surgeLatitude}(e)) to -60$^{\circ}$ (Figure \ref{fig:surgeLatitude}(f)).  With a strength larger than 3 G, the surge is significantly stronger than following surges, of which the strength does not exceed 1G.  For intermediate latitudes that are not close to the pole (where the net flux does not pile up), like latitudes -35$^{\circ}$ (Figure \ref{fig:surgeLatitude}(a)) and -45$^{\circ}$ (Figure \ref{fig:surgeLatitude}(b)), we can see from the black observational curves that the surges after the prominent surge we consider are of different signs, so their influence on the final polar field is fairly small.  As shown by the red solid curves, our simulation of the surge of interest maintains its key features, while being not as narrow as the observational surge.  The simulated surge at 35$^{\circ}$ is located somewhat away from the center of the observational surge, which is possibly a result of AR assimilation random error.  The initial difference is, however, reduced by the transport of flux.  The magnetic field evolution in the simulation after CR2159, that is, after all AR assimilation, stays close to that of observations.  This is similar to the situation of polar field evolution.  Thus we can conclude that the surge we consider defines the major evolution of magnetic field of the southern hemisphere since CR2145.

The surge is primarily generated by ARs within ACs.  To show this, we run a simulation with only ARs within the two ACs described in Subsection \ref{subsec:identifyResults} and CR2144 as the initial field.  The generated surge by only ARs in the two ACs is shown as the red dash-dot curves in Figure \ref{fig:surgeLatitude}.  The resulting surge is fairly close to the overall result of the surge where all ARs are assimilated, with matching surge strength and width.  The ARs in the two ACs generate the prominent surge, and in turn have a long-lasting influence on the southern polar field during the latter half of cycle 24.


\subsection{Evolution characteristics of ARs within ACs} \label{subsec:differentARs}



ACs are characterized as consistent flux emergence as well as cancellation, which affects the configuration of surface flux.  In order to discuss how this feature is simulated in the SFT model, we choose a sequence of ARs between 180$^{\circ}$ and 270$^{\circ}$ longitudes during CRs 2152-2153 of the southern hemisphere to simulate their evolution together.  The parameters of these ARs are shown in Table \ref{tab:acarlist}.  Most ARs listed in Table \ref{tab:acarlist} have large and positive tilt angles, but with a wide range of variation.  This is expected since statistical studies of tilt angles have shown that tilt angles of ARs have a large deviation from the Joy's law \citep{1919ApJ....49..153H}.  The scatter of the tilts is regarded as the result of the buffeting by convective turbulence during the rise of flux tubes through the convection zone to form ARs \citep{2013SoPh..287..239W}.  The poleward flux originates from a part of the trailing polarity depending on the tilt angles and latitudes of ARs in the AC.  We note that, among the listed ARs, some are too close to be identified individually.  Hence some identified regions may contain more than one NOAA AR, e.g., the pair of ARs 12104 and 12107. We start the simulation without initial field and assimilate the ARs on their corresponding day.  The simulation is run for 10 years for the field to reach their finial state.


\begin{deluxetable*}{cccccccc}[b!]
\tablecaption{Parameters of the selected ARs within the AC \label{tab:acarlist}}
\tablecolumns{8}
\tablenum{3}
\tablewidth{0pt}
\tablehead{
\colhead{Day of emergence\tablenotemark{a}} &
\colhead{Latitude} &
\colhead{Longitude} &
\colhead{Area} &
\colhead{Positive flux} &
\colhead{Negative flux} &
\colhead{Tilt} &
\colhead{NOAA/AR number}\\
\colhead{} & \colhead{(degree)} & \colhead{(degree)} &
\colhead{($\mu$Hem)} & \colhead{(10$^{20}$Mx)} & \colhead{(10$^{20}$Mx)} & \colhead{(degree)} & \colhead{}
}
\startdata
195 & -15.2 & 262.9 & 4730.8 & 238.2 & -154.6 & 47.0 & 12104,12107 \\
197 & -8.5 & 237.8 & 3386.5 & 151.1 & -179.9 & 1.8 & 12108,12110 \\
199 & -8.1 & 218.9 & 2845.9 & 162.8 & -128.4 & 5.0 & 12109 \\
224 & -8.5 & 249.5 & 2198.8 & 88.8 & -51.1 & 43.6 & 12127 \\
225 & -18.5 & 238.5 & 1151.5 & 35.3 & -33.4 & 7.2 & 12131 \\
227 & -20.2  & 209.4   &  2368.9    &   84.5   &   -94.4  &    25.8   &   12132	\\
\enddata
\tablenotetext{a}{Since the end of CR2144 (2013 Dec 19).}
\end{deluxetable*}

\begin{figure}
\gridline{\fig{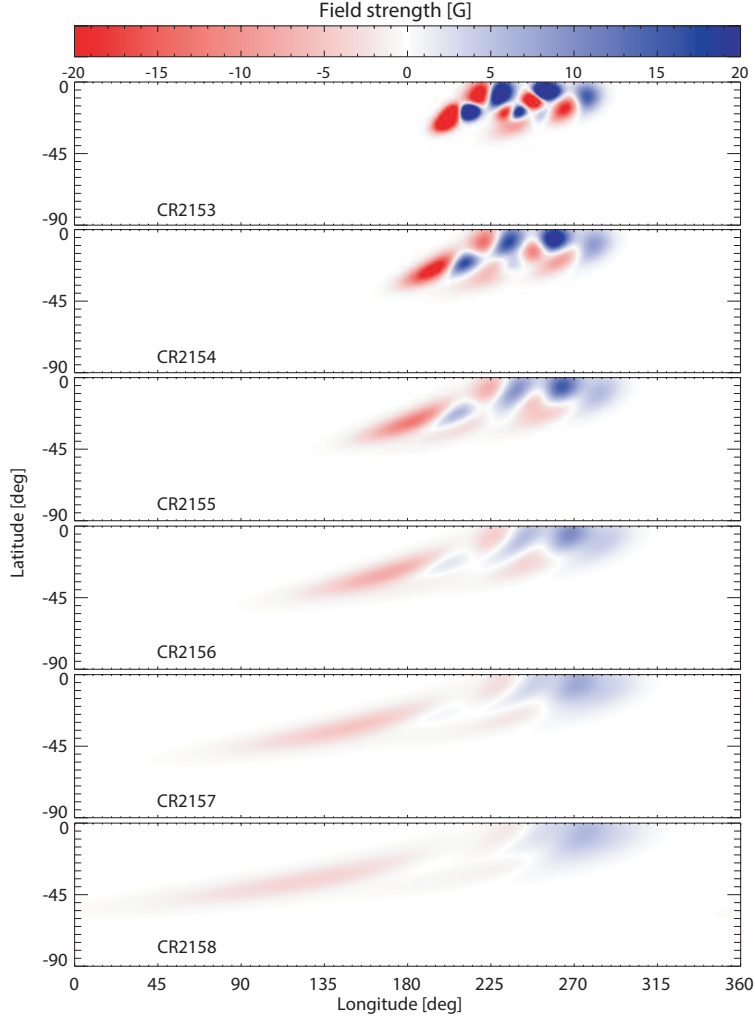}{0.55\textwidth}{}
          }
\caption{Stack plot created from the southern hemisphere of simulated synoptic maps of ARs assimilated in Subsection \ref{subsec:differentARs} from CR2153 to CR2158.  Each simulated synoptic map is displayed in equal space in latitude.\label{fig:simAC}}
\end{figure}

During the simulation, opposite polarities of close ARs cancel, which is also seen in observations.  Figure \ref{fig:simAC} shows the simulated synoptic maps for CRs 2153-2158.  In the simulation, a large area of concentrated flux of the leading polarity, that is, positive polarity is formed at the edge of the AC, as the flux from ARs cancels out.  Apparently the polarities of the same sign from different ARs tend to merge into larger area of flux as a result of cancellation between ARs, similar to the observations of \citet{2008ApJ...686.1432G}.  Eventually a long-lived unipolar region is formed, and affects nearby ARs, as described in Subsection \ref{sibsec:identifyARs}.  The unipolar region of positive polarity still resides at low to intermediate latitudes after CR2156 for several rotations.  Hence it exists close to AR12192 as it evolves for a considerably long time.  The extended coexistence of the unipolar region and the diffusive AR12192 affects the identification of AR12192 and new weak flux emergence, as well as the analysis of the development of AR12192 in the following CRs.  In the SFT model, ARs interact only in the form of cancellation from supergranular diffusion.  Consequently, the observed process of gathering same sign polarities and the formation of unipolar regions is mostly a result of flux cancellation.  Such cancellation of ARs is common for ACs, as described in Subsection \ref{subsec:identifyResults}, and in the observations of \citet{2001ApJ...558..888G} and \citet{2008ApJ...686.1432G}.


We present the simulated magnetic butterfly diagram shown in Figure \ref{fig:simMagbfly}(c), for assimilating only a part of the ACs.  The butterfly diagram shows that the part of ACs assimilated creates a section of the poleward surge of the trailing polarity.  As for the leading polarity, both poleward migration and transequatorial migration occur.  This is the typical evolution for a single AR at intermediate emerging latitudes.  After the surge, the large unipolar region of the positive polarity migrates to both poles, which weakens the southern polar field and strengthens the northern polar field.  The features of the generated surge can be shown by the red dashed curves in Figure \ref{fig:surgeLatitude}.  According to the peaks of the curves, this surge has a relatively stable strength of 0.7-0.8 G, which is roughly a fourth to a third of the overall surge strength.


\section{Discussion and conclusion} \label{sec:conclusion}

We have applied a data-driven SFT model to simulate the formation and evolution of the prominent poleward surge originating from ARs during CRs 2145-2159 on the southern hemisphere in solar cycle 24.  We find the strength and shape of the surge is primarily determined by ARs within the two ACs, which consists of about half the number of all ARs identified.  Our simulations show that the ARs within the ACs generating the surge have a strong long-term effect on the final polar field at cycle minimum, while other surges during the latter half of cycle 24 only have considerably less effect on the final polar field.  The development of the southern polar field of the latter half of cycle 24 is predominantly shaped by ARs generating the surge.  Without the ARs within the ACs generating the prominent surge, the final polar field would be less than 1 G, and even reverse to previous polarities.  The ARs within the ACs effectively build up the polar field of cycle 24 minimum, thus determining the strength of cycle 25 in the framework of B-L-type dynamos, instead of entering a next Maunder minimum.  The surge we consider is different from the surge studied by \citet{2015SoPh..290.3189Y}.  Our surge is stronger and has longer persistence than the surge studied by \citet{2015SoPh..290.3189Y}.  The ARs generating our surge mostly lie between -10$^{\circ}$ to -20$^{\circ}$ latitudes, which are lower than that of \citet{2015SoPh..290.3189Y}'s, and our surge has significantly larger long-term polar field influence.  This is consistent with the relationship between ARs' final polar field contributions and emerging latitudes proposed by \citet{2014ApJ...791....5J}.  Since the strength and long-term polar field influence of the surge is unique to cycle 24, it makes sense to refer to it as a ``super surge'', similar to the idea of super ARs.  This feature is a notable supplementary to the AC-surge-polar field relationship obtained from previous observations.



In SFT simulations, ARs within ACs are similar to a single, large AR in terms of surge generation and polar field influence, as the observed process of ARs' emergence and cancellation creating large regions of leading and trailing polarities can be simulated by SFT processes.  For large ARs with large latitude coverage, the long-term polar field influence is focused on the flux located around lower latitudes.  Thus it may deviate from estimations given by overall AR parameters.  The concept was first proposed by \citet{2019ApJ...871...16J} based on the evolution of a $\beta\gamma\delta$ type AR.  The effect of size asymmetry for two polarities of BMR type ARs examined in \citet{2019ApJ...883...24I} is also consistent with this concept.  ACs and ARs with large latitude coverage can contribute to both surge and polar field, and it is likely that the surge is not followed by prominent surge of the other polarity.  Such a surge can be regarded as the cause of long-term polar field change. 

Even for surges causing long-term polar field change, the contributed magnetic flux to the final polar field consists of only a very small part of originating ARs' total flux.  The percentage depends on the tilt angle and latitude of the ARs.  The AR flux decreases monotonically as a result of diffusive cancellation at the neutral line.  Poleward fluxes from separate ARs still cancel in the polar caps if they have different polarities. The axial dipole field, which results from AR tilt angles, has the longest lifetime among different orders of multipoles \citep{2000GeoRL..27..621W,2006A&A...446..307B}.  It can survive the SFT process at lower latitudes and contribute to the polar field.  The final polar field of a cycle is determined by the total axial dipole field of ARs, including those in ACs, emerged during the cycle, if the initial field and the transport parameters are fixed.


ACs are abundant with intense flux emergence and cancellation.  The spatial and temporal proximity of ARs in ACs is a potential cause of an intrinsic problem of automatically identifying and precisely quantifying ARs in ACs.  The decaying ARs are likely to be mixed with new emerging flux as well as unipolar regions from previous ARs in ACs.  Meanwhile, new flux emergence and associated flux cancellation can alter the configuration of ARs.  ACs are also considered to bear a collective magnetic structure unable to reduce to single ARs \citep{1983IGAFS..65..129B}.  All these contribute to the difficulty of automatically identifying and assimilating ARs within ACs.  For the formation of unipolar regions in simulations, we note that flux cancellation in SFT simulations is rather simple, not completely realistic.  The removal of flux on the solar surface includes small-scale diffusions and large-scale retraction of flux and U-loops \citep{1992ASPC...27...89V}.  The importance of large-scale retraction of flux and U-loops on removal of flux still needs to be examined.

Our simulations do not include inflows toward activity belts, but these are expected to take part in ACs.  The strength of inflows is dependent on the flux concentration in activity belts due to temperature deficit \citep{2003SoPh..213....1S,2008SoPh..251..241G,2010ApJ...720.1030C}.  So it affects ACs, as ACs' emergence is related to the cycle strength.  A detailed examination of inflows on ACs is needed for future SFT simulations.

Our assimilation method for ARs with real configurations is able to produce reliable poleward surge and polar field evolution.  However, its accuracy for single ARs is still limited.  The two polarities of identified ARs are not balanced, and this imbalance is not totally eliminated by our procedure of enlarging the weaker polarity.  Observationally ARs are not balanced in positive and negative polarity, as the trailing polarity may be more diffusive so that some flux may remain undetected by observations.  Meanwhile, ARs' magnetic field may be connected to other magnetic features.  Besides these observational limitations, resizing magnetograms might also be possible source of extra imbalance.  This imbalance is required to be dealt with in future simulations.

\acknowledgments

We thank the referee for the valuable comments and suggestions on improving the manuscript.  The SDO/HMI data are courtesy of NASA and the SDO/HMI team.  This research was supported by the National Natural Science Foundation of China through grant Nos. 11873023, 11322329, and 11533008, Key Research Program of Frontier Sciences of CAS through grant No. ZDBS-LY-SLH013, and the B-type Strategic Priority Program of CAS through grant No. XDB41000000.  J.J. acknowledges the International Space Science Institute Teams 474 and 475.

\appendix

\section{Validating the SFT simulation results} \label{sec:validateSFT}

In order to examine the reliability of the SFT simulation results, we introduce two methods of validating the SFT simulations.

One method is to compare the one year evolution of the simulated polar field with observations.  If the initial field and the transport terms were ideal, the SFT simulated polar field would be comparable to observations for approximately a year.  This is because that the variation of the polar field is due to the magnetic flux originated from the activity belt.  It takes a few years for the flux to be transported to the poles. The red and blue lines in Figure \ref{fig:timePfAll}, which are represented in Appendix as Figure \ref{fig:timePfAll2}, are results of the simulations starting from CR2144 with and without ARs assimilated, respectively.  In both cases, we find that they are consistent with observational data from HMI synoptic maps (black lines) for about 1 year since CR2145.  Apart from these simulations described already in Subsection \ref{subsec:surge}, we introduce another simulation with the initial field set as CR2159 and without ARs assimilated (orange lines in Figure \ref{fig:timePfAll2}). Its polar field evolutions also fit the standard of the method. These results show that the transport terms in the SFT model are generally reliable.

The other validation is to compare two SFT simulations starting at different CRs while ending at the same CR.  Still if the initial field and the transport terms were ideal, the two SFT simulations would strictly follow each other during the overlapped time period. Here we compare the two simulations starting at CR2144 and CR2159, respectively.  We verify the consistency of the surface field's development from CR2159 to the end of the simulations.  We find from Figure \ref{fig:timePfAll2} that the red lines (starting at CR2144) and the orange lines (starting at CR2159) are fairly comparable.  This result further supports the reliability of the transport terms in the SFT model.  The result also shows that the AR identification and assimilation methods that we apply during CRs 2145-2159 produce source terms that are close to the realistic flux emergence. The results of the two methods confirm the reliability of our simulation results.

\begin{figure}
\gridline{\fig{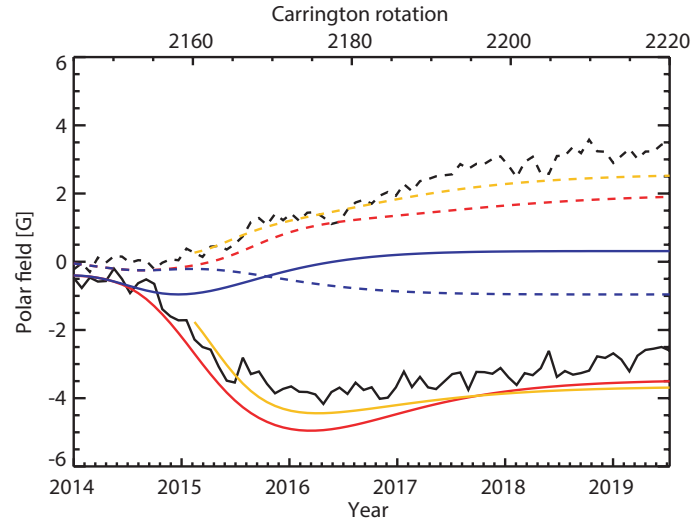}{0.5\textwidth}{}}
\caption{Validations of the SFT simulations.  Black, red, and blue lines are the same as that in Figure \ref{fig:timePfAll}, but for comparisons of the first-year evolution between simulations (red and blue lines) and observations (black lines).  The orange lines represent the results of the simulation with initial field CR2159 and without ARs assimilated.\label{fig:timePfAll2}}
\end{figure}

\end{document}